\documentclass[amsmath,amssymb,10pt,eqsecnum]{revtex4}
\usepackage{hyperref, graphicx, amsmath, amsmath, amssymb}

\newcommand{\virt}[0]{\hat{\delta}}

\newcommand{\half}[0]{\frac{1}{2}}

\newcommand{\ld}[0]{\mathcal{L}}

\newcommand{\dd}[0]{\textrm{d}}
\newcommand{\defn}[0]{\equiv}

\newcommand{\qsubrm}[2]{{#1}_{\scriptscriptstyle{\textrm{#2}}}}

\newcommand{\subsm}[2]{{#1}_{\scriptscriptstyle{#2}}}
\newcommand{\supsm}[2]{{#1}^{\scriptscriptstyle{#2}}}

\newcommand{\AW}[0]{A_{\mathcal{W}}}
\newcommand{\BW}[0]{B_{\mathcal{W}}}
\newcommand{\CW}[0]{C_{\mathcal{W}}}
\newcommand{\DW}[0]{D_{\mathcal{W}}}
\newcommand{\EW}[0]{E_{\mathcal{W}}}

\newcommand{\sol}[0]{\ld_{\scriptscriptstyle\{2\}}}

\newcommand{\hct}[0]{\mathcal{H}}
\renewcommand{\figurename}{Figure}
\newcommand{\ep}[0]{{ {\delta}_{\scriptscriptstyle{\rm{E}}}}}

\def\be{\begin{equation}}
\def\ee{\end{equation}}
\def\bea{\begin{eqnarray}}
\def\eea{\end{eqnarray}}
\def\bse{\begin{subequations}}
\def\ese{\end{subequations}}

\newcommand{\spnab}[0]{{\overline{\nabla}}{}}

\newcommand{\fref}[1]{{Figure \ref{#1}}}

\let\oldsqrt\sqrt
\def\sqrt{\mathpalette\DHLhksqrt}
\def\DHLhksqrt#1#2{%
\setbox0=\hbox{$#1\oldsqrt{#2\,}$}\dimen0=\ht0
\advance\dimen0-0.2\ht0
\setbox2=\hbox{\vrule height\ht0 depth -\dimen0}%
{\box0\lower0.4pt\box2}}

\begin{document}

\title{Massive gravity, the elasticity of space-time and perturbations in the dark sector}
\author{Richard A. Battye}
\email{richard.battye@manchester.ac.uk}
\affiliation{Jodrell Bank Centre for Astrophysics, School of Physics and Astronomy, The University of Manchester, Manchester M13 9PL, U.K}
\author{Jonathan A. Pearson}
\email{jonathan.pearson@durham.ac.uk}
\affiliation{Department of Mathematical Sciences, Durham University, South Road, Durham, DH1 3LE, U.K}
\affiliation{Jodrell Bank Centre for Astrophysics, School of Physics and Astronomy, The University of Manchester, Manchester M13 9PL, U.K}

\date{\today}

\begin{abstract}We consider a class of modified gravity models where the terms added to the standard Einstein-Hilbert Lagrangian are just a function of the metric only. For linearized perturbations around an isotropic space-time, this class of models is entirely specified by a rank 4 tensor that encodes possibly time-dependent masses for the gravitons. This tensor has the same symmetries as an elasticity tensor, suggesting an interpretation of massive gravity as an effective rigidity of space-time. If we choose a form for this tensor which is compatible with the symmetries of FRW and  enforce full reparameterization invariance, then the only theory possible is a cosmological constant. However, in the case where the theory is only time translation invariant, the ghost-free  massive gravity theory is equivalent to the elastic dark energy scenario with the extra Lorentz violating vector giving rise to 2 transverse and 1 longitudinal degrees of freedom, whereas when one demands spatial translation invariance one is left with scalar field theory with a non-standard kinetic term. \end{abstract}
 
\maketitle
\section{Introduction}
The realization that the Universe appears to be accelerating has fuelled the search to alternative theories of gravity \cite{Clifton:2011jh} as a possible explanation for what has become called \textit{the dark sector}. In this paper we will focus on what is the simplest subset of such theories, in which  the dark sector Lagrangian is only a function of the metric (and no extra derivatives thereof) following the approach discussed in \cite{Battye:2012eu}. The action for this type of theory is given by
\bea
\label{eq:sec:intro-action}
S = \int \dd^4x\sqrt{-g}\, \bigg[ R + 16\pi G \qsubrm{\ld}{m} - 2 \qsubrm{\ld}{d}(g_{\mu\nu})\bigg].
\eea
If $T_{\mu\nu}$ is the energy-momentum tensor for the matter sector Lagrangian $\qsubrm{\ld}{m}$ and $U_{\mu\nu}$ is that associated with the dark sector Lagrangian $\qsubrm{\ld}{d}$, then the Einstein equation is $G_{\mu\nu}=8\pi GT_{\mu\nu}+U_{\mu\nu}$. Typically, we will be interested in spacetimes which are isotropic where $U_{\mu\nu} = \rho u_{\mu}u_{\nu} + P\gamma_{\mu\nu}$ can specified in terms of a density, $\rho$, and pressure, $P=w\rho$. There are two classes of theories  which can be described by (\ref{eq:sec:intro-action}): elastic dark energy and theories of massive gravitons. Typically, theories of massive gravitons have been considered as a fundamental theory around Minkowski space-time, but it is also possible to think about masses for the gravitons as being induced by a some unknown effective physics encoded by $\qsubrm{\ld}{d}$.

The study of massive gravity theories (see, for example, \cite{Rubakov:2008nh, Hinterbichler:2011tt}) has a long history. This started with the linearized theories of Pauli and Feirz \cite{Fierz:1939ix}, progressing to the studies of Boulware and Deser \cite{Boulware:1973my}. It has received a new lease of life in recent times with the proposal of non-linear dRGT massive gravity theory \cite{deRham:2010ik, deRham:2010kj, Hassan:2011hr, Hassan:2011vm} and its connections to the Vainshtein screening mechanism \cite{18084, Babichev:2010jd, DeFelice:2011th, Sbisa:2012zk}.  Massive gravity theories are built upon the pretext that the resulting theory should be ``ghost-free'' \cite{Creminelli:2005qk, Alberte:2010qb, Alberte:2010it, deRham:2011rn, Hassan:2011hr, Hassan:2011tf, Hassan:2011zd, Paulos:2012xe}, and have begun to be studied in cosmological backgrounds \cite{Grisa:2009yy, Berkhahn:2010hc, Berkhahn:2011hb, Berkhahn:2011jh, D'Amico:2011jj, Gumrukcuoglu:2011ew, Gumrukcuoglu:2011zh, Crisostomi:2012db, Gratia:2012wt, deFelice1206.2080, Volkov:2011an, Volkov:2012zb, D'Amico:2012pi}. Such theories are usually presumed to be Lorentz invariant which leads to the Pauli-Fierz tuning.  Giving up Lorentz-invariance is another way to remove ghosts from the theory, as pointed out by \cite{ArkaniHamed:2003uy, Rubakov:2004eb} and further studied in \cite{Dubovsky:2004sg, Gabadadze:2004iv, Rubakov:2008nh, Blas:2009my, Comelli:2012vz}.
 
Elastic dark energy (EDE) is an idea which has been developed from relativistic elasticity theory \cite{BF01645505, carterqunt_1977, Carter26081980, PhysRevD.7.1590, Carter21111972, Carter:1982xm, 1978ApJ2221119F}. The basic concept is that the stress-energy component  responsible for the dark energy has rigidity which stabililizes perturbations that would, if modelled as those of a perfect fluid, give rise to exponential growth in the density contrast. The framework was adapted for cosmological purposes in \cite{PhysRevD.60.043505,Battye:1999eq} in order to provide a phenomenological model for domain wall  as an explanation for accelerated expansion which have  $P/\rho=\qsubrm{w}{dw} = - 2/3$. However, in principle the equation of state parameter,  $w$,  is allowed to take ``any'' value  so long as the rigidity modulus, $\mu$, is sufficiently large. Indeed, the theory is well-defined in the limit $w\rightarrow -1$ where the elastic medium becomes a ``cosmological constant'' and $w\rightarrow 0$ and $\mu\rightarrow 0$ which corresponds to cold dark matter. The standard assumption,  which we will use in this paper, is that the elasticity tensor is isotropic, but one can also construct anisotropic models \cite{Battye:2006mb, Battye:2009ze}.

The aim of this paper is to point out the connections between linearized massive gravity theories, EDE and the framework for linearized perturbations in the generalized models for the dark sector discussed in \cite{Battye:2012eu, Pearson:2012kb}. The reason for this connection is that, at linearized order, one can represent all possible Lagrangians by a generalized function which is quadratic in the fields. In the specific case we are concerned with here, this is just a quadratic function of the metric which is parameterized by a rank-4 tensor. This has the same symmetries as an elasticity tensor, suggesting an interpretation of massive gravitons as creating an effective rigidity of spacetime.  This tensor can split in a way which is compatible with the symmetries of the FRW spacetime, that is more general than the usual Pauli-Fierz case.  We will make a survey of the possible mechanisms by which ghost modes can be removed, both Lorentz invariant and violating. The ghosts are associated with a breaking of reparameterization invariance and we show that its re-imposition leads to 3 interesting sub-cases, one of which is compatible with Lorentz invariance which is a cosmological constant and the other two which violate either  time or spatial translation invariance. The one which violates spatial translation invariance happens to be the EDE model which we see is a Lorentz-violating  ghost-free massive gravity theory.

\section{The general ``metric only'' theory}
The action which will give linearized field equations for the perturbed field variables is given by
\bea
\label{eq:s2_action}
\subsm{S}{\{2\}} = \int \dd^4x\sqrt{-g} \, \bigg[ \Diamond^2R + 16 \pi G \Diamond^2\qsubrm{\ld}{m} - 2 \sol \bigg].
\eea
We use $\Diamond^2Q$ to denote the second measure-weighted variation of the quantity $Q$, defined as $\Diamond^2Q \defn \frac{1}{\sqrt{-g}}\delta^2(\sqrt{-g}Q)$.  $\sol$ is the Lagrangian for dark sector perturbations, given by
\bea
\label{eq:sec:sol-mass-H-W}
\sol =\frac{1}{8} \mathcal{W}^{\mu\nu\alpha\beta} \delta g_{\mu\nu}\delta g_{\alpha\beta},
\eea
where $\delta g_{\mu\nu}$ is the metric fluctuation. This clearly looks like a mass term for the metric perturbations. The tensor 
\bea
\label{eq:Sec:mass-matrx-symms}
\mathcal{W}^{\mu\nu\alpha\beta}=\mathcal{W}^{(\mu\nu)(\alpha\beta)}=\mathcal{W}^{\alpha\beta\mu\nu}
\eea
is the \textit{mass matrix} determining how the components of $\delta g_{\mu\nu}$ mix to provide the mass; all linearized massive gravities are encoded by choices of $\mathcal{W}$.  Hence, the complete linearized theory we study is
\bea
\label{eq:sec:pert-met-only-ful_theory}
\subsm{S}{\{2\}} = \int \dd^4x\sqrt{-g} \, \bigg[\Diamond^2R + 16 \pi G \Diamond^2\qsubrm{\ld}{m}  - \tfrac{1}{4} \mathcal{W}^{\mu\nu\alpha\beta} \delta g_{\mu\nu}\delta g_{\alpha\beta} \bigg].
\eea
This   theory contains metric fluctuations which have a kinetic term, and a mass term. We will see that the spatial components of $\mathcal{W}$ can be interpreted as an elasticity tensor.

To isolate the degrees of freedom in the theory, $\delta g_{\mu\nu}$ can be decomposed as
\bea
\label{eq:sec;decomps-xi-whole-dg-h-xi}
\delta g_{\mu\nu} = h_{\mu\nu} + 2\nabla_{(\mu}\xi_{\nu)}.
\eea
In the parlance of \cite{carterqunt_1977, 1978ApJ2221119F, Battye:1995hv, Battye:1998zk} $h_{\mu\nu}$ is the Eulerian metric perturbation and $\xi_{\mu}$ is a vector field representing possible coordinate transformations. In standard General Relativity, the action is independent of $\xi_{\mu}$, but in more general theories this can become a physical field. This formulation is equivalent to what is sometimes called the Stuckelberg trick \cite{ArkaniHamed:2002sp, Hinterbichler:2011tt, deRham:2011rn}. Inserting (\ref{eq:sec;decomps-xi-whole-dg-h-xi}) into the action (\ref{eq:sec:pert-met-only-ful_theory}) and integrating by parts reveals that
\bea
\label{eq:action_s2_cons}
\qsubrm{S}{\{2\}} &=&  \int \dd^4x\sqrt{-g} \, \bigg[ \big(\ep G^{\mu\nu} - 8 \pi G \ep T^{\mu\nu} - \ep U^{\mu\nu}\big)h_{\mu\nu}   +2 \xi_{(\mu}\ep(\nabla_{\nu)} U^{\mu\nu}) \bigg],
\eea
where the variational operator ``$\ep$'' denotes that the quantity is evaluated with the metric perturbation variable $h_{\mu\nu}$ (rather than $\delta g_{\mu\nu}$), and where we have defined the perturbed dark energy-momentum tensor
\bea
\label{eq:sec:per-dark-emt-fdkjhfj}
\ep U^{\mu\nu} = - \tfrac{1}{2} \big( \mathcal{W}^{\mu\nu\alpha\beta} + U^{\mu\nu}g^{\alpha\beta} \big) \delta g_{\alpha\beta}- \xi^{\alpha}\nabla_{\alpha}U^{\mu\nu} + 2U^{\alpha(\mu}\nabla_{\alpha}\xi^{\nu)}.
\eea
It is now a simple matter to obtain the functional derivatives of the action with respect to the perturbed metric $h_{\mu\nu}$ and $\xi^{\mu}$-fields,
\bse
\bea
\frac{\virt}{\virt h_{\mu\nu}} \qsubrm{S}{\{2\}}= \ep G^{\mu\nu} - 8 \pi G \ep T^{\mu\nu} - \ep U^{\mu\nu}=0,
\eea
\bea
\label{eq:sec:action-vzry-xi}
\frac{\virt}{\virt \xi_{\mu}} \qsubrm{S}{\{2\}} = \ep(\nabla_{\nu} U^{\mu\nu})=0.
\eea
\ese
The variational principle was used to demand that these expressions vanish, yielding the perturbed gravitational field equations and perturbed conservation equation respectively.
%It is useful to see that the contribution from the  $\xi^{\mu}$-field to the action (\ref{eq:action_s2_cons}) can be written as
%\bea
%\qsubrm{S}{\{2\}} &\supset& \int \dd^4x\sqrt{-g}  \xi_{(\mu}\ep(\nabla_{\nu)} U^{\mu\nu})  \nonumber\\
%&=& \int \dd^4x\sqrt{-g} \bigg\{ \xi_{\nu}L^{\mu\nu\alpha\beta}\nabla_{\mu}\nabla_{\alpha}\xi_{\beta} + \xi_{\nu}(\nabla_{\mu}\mathcal{W}^{\mu\nu\alpha\beta}) \nabla_{\alpha}\xi_{\beta}  +\xi_{\nu}(\nabla_{\mu}\nabla_{\alpha}U^{\mu\nu}) \xi^{\alpha} \nonumber\\
%&&\qquad\qquad\qquad+ \tfrac{1}{2} \xi_{\nu}\big((\nabla_{\mu}\mathcal{W}^{\mu\nu\alpha\beta}) h_{\alpha\beta}+C^{\mu\nu\alpha\beta}  \nabla_{\mu}h_{\alpha\beta} \big)\bigg\}.
%\eea
% 
Using (\ref{eq:sec:per-dark-emt-fdkjhfj}) to evaluate (\ref{eq:sec:action-vzry-xi}) yields
\bea
\label{eq:sec:xi-eom-vector}
L^{\mu\nu\alpha\beta}\nabla_{\mu}\nabla_{\alpha}\xi_{\beta} + (\nabla_{\mu}\mathcal{W}^{\mu\nu\alpha\beta}) \nabla_{\alpha}\xi_{\beta}  +(\nabla_{\mu}\nabla_{\alpha}U^{\mu\nu}) \xi^{\alpha}  =- \tfrac{1}{2} \big((\nabla_{\mu}\mathcal{W}^{\mu\nu\alpha\beta}) h_{\alpha\beta}+P^{\mu\nu\alpha\beta}  \nabla_{\mu}h_{\alpha\beta} \big),
\eea
where we defined the \textit{effective metric} $L^{\mu\nu\alpha\beta}$ and \textit{derivative-coupling} $P^{\mu\nu\alpha\beta}$ terms,
\bea
L^{\mu\nu\alpha\beta} \defn   \mathcal{W}^{\mu\nu\alpha\beta} +  U^{\mu\nu}g^{\alpha\beta} - 2U^{\alpha(\mu}g^{\nu)\beta} ,
\qquad
P^{\mu\nu\alpha\beta} \defn  \mathcal{W}^{\mu\nu\alpha\beta}+U^{\alpha\beta}g^{\mu\nu} - 2 g^{\nu\beta}U^{\alpha\mu} .
\eea

We   impose spatial isotropy upon the background  with the   (3+1) decomposition and in doing so we will  obtain the most general linearized massive gravity Lagrangian compatible with spatial isotropy of the background.  We   foliate the 4D spacetime by 3D sheets with a time-like unit vector, $u_{\mu}$, being everywhere orthogonal to the sheets. The 4D spacetime has metric $g_{\mu\nu}$, and the 3D sheets have metric $\gamma_{\mu\nu}$. The (3+1) decomposition of the 4D metric is $g_{\mu\nu} = \gamma_{\mu\nu} - u_{\mu}u_{\nu}$, where $u^{\mu}u_{\mu}=-1$,  $\gamma^{\mu\nu} u_{\mu}=0$. This structure provides  an extrinsic curvature tensor $K_{\mu\nu}=K_{(\mu\nu)}$ on the 3D sheets, given by $K_{\mu\nu} = \nabla_{\mu}u_{\nu}$ and satisfying   $u^{\mu}K_{\mu\nu}=0$ (the extrinsic curvature tensor is given by $K_{\mu\nu} = \frac{1}{3}K\gamma_{\mu\nu}$). We   define ``time''  and ``space'' differentiation as the derivative operator projected along   the time and space directions,
\bea
\dot{\psi} \defn u^{\mu}\nabla_{\mu}\psi,\qquad \spnab_{\mu}\psi \defn {\gamma^{\nu}}_{\mu}\nabla_{\nu}\psi.
\eea
Using this technology we   decompose the gradient of a scalar into two orthogonal terms,
\bea
\nabla_{\mu}\psi = - \dot{\psi} u_{\mu} + \spnab_{\mu}\psi.
\eea
This enables us to find the values of two useful ``kinetic scalars'',
\bse
\label{eq:sec:31_kin_scalars}
\bea
\nabla^{\mu}\psi\nabla_{\mu}\psi =- \dot{\psi}^2 + \spnab^{\mu}\psi\spnab_{\mu}\psi,
\eea
\bea
\square\psi \defn \nabla^{\mu}\nabla_{\mu}\psi = -\ddot{\psi} +\spnab^{\mu}\spnab_{\mu}\psi.
\eea
\ese
The last term of each expression simply selects out the spatial derivatives of the scalar field.  Another useful application of the (3+1) decomposition is to find all the freedom in a tensor which is compatible with spatial isotropy of the background spacetime.

We use the (3+1) decomposition to isolate the components of the   perturbed metrics   by writing
\bse
\label{eq:decomp-H-nudhfkdh-sec-tot}
\bea
\label{eq:decomp-H-nudhfkdh-sec}
\delta g_{\mu\nu} = 2 \Phi u_{\mu}u_{\nu} + 2N_{(\mu}u_{\nu)} + \bar{H}_{\alpha\beta} {\gamma^{\alpha}}_{\mu}{\gamma^{\beta}}_{\nu},
\eea
\bea
h_{\mu\nu} = 2 \phi u_{\mu}u_{\nu} + 2n_{(\mu}u_{\nu)} + \bar{h}_{\alpha\beta} {\gamma^{\alpha}}_{\mu}{\gamma^{\beta}}_{\nu},
\eea
and we isolate the time-like and space-like components of the vector field via
\bea
\xi_{\mu} = - \chi u_{\mu} +  {\omega}_{\mu},
\eea
\ese
where $N^{\mu}u_{\mu}=n^{\mu}u_{\mu}=0, u^{\mu}\bar{H}_{\mu\nu} =u^{\mu}\bar{h}_{\mu\nu}=0$ and $u^{\mu}{\omega}_{\mu}=0$.

In \cite{Battye:2012eu} we showed that the general decomposition of the mass-matrix $\mathcal{W}$ compatible with spatial isotropy is
\bea
\label{eq:Sec:31_w}
\mathcal{W}^{\mu\nu\alpha\beta} &=& \AW u^{\mu}u^{\nu}u^{\alpha}u^{\beta} + \BW\big( u^{\mu}u^{\nu}\gamma^{\alpha\beta} + u^{\alpha}u^{\beta}\gamma^{\mu\nu}\big)  + 2\CW\big( \gamma^{\alpha(\mu}u^{\nu)}u^{\beta} + \gamma^{\beta(\mu}u^{\nu)}u^{\alpha}\big)\nonumber\\
&&\qquad\qquad + \DW \gamma^{\mu\nu}\gamma^{\alpha\beta} + 2\EW \gamma^{\mu(\alpha}\gamma^{\beta)\nu},
\eea
where there are only 5 free functions which only depend on time. 
Using the mass matrix (\ref{eq:Sec:31_w}) and  (\ref{eq:decomp-H-nudhfkdh-sec})  in  the Lagrangian (\ref{eq:sec:sol-mass-H-W})     yields
\bea
\label{eq:sec:sol-lag-met-only-laper-a}
8\sol &=& 4\AW \Phi^2 + 4\BW\Phi \bar{H} + 2 \CW N_{\alpha}N^{\alpha}  + \DW \bar{H}^2 + 2 \EW \bar{H}^{\alpha\beta}\bar{H}_{\alpha\beta}.
\eea
This can be written in terms of ``graviton masses'' (see e.g., \cite{Rubakov:2008nh})  where the free functions $\{\AW,\ldots, \EW\}$ are given by
\bea
2\sol &=& m_0^2( \delta g_{00})^2 + 2 m_1^2(\delta g_{0i})^2 - m_2^2(\delta g_{ij})^2 + m_3^2(\delta g_{ii})^2 - 2 m_4^2\delta g_{00} \delta g_{ii}
\eea
with $\AW = m_0^2$,  $\BW = -2m_4^2$, $\CW = 4m_1^2$, $\DW = 4m_3^2$ and $\EW = -2m_2^2$.
One should keep in mind, therefore, that when we talk about the $\{\AW,\ldots, \EW\}$ we are actually talking about the graviton masses $m_i^2$, albeit ones that depend on time. The values of these masses for a ``Goldstone''   theory are given in \cite{Dubovsky:2005dw} and   those which are induced by perturbations in scalar fields in \cite{Battye:2012eu}.

Using (\ref{eq:decomp-H-nudhfkdh-sec-tot}) to evaluate (\ref{eq:sec;decomps-xi-whole-dg-h-xi}) yields
\bea
\label{eq:sec:sol-lag-met-only-laper-ab}
\Phi = \phi + \dot{\chi},
\qquad N_{\alpha}= n_{\alpha} - \dot{\varpi}_{\alpha} - \spnab_{\alpha}\chi,
\qquad
\bar{H}_{\alpha\beta} = \bar{h}_{\alpha\beta} + 2\spnab_{(\alpha}\omega_{\beta)}  - \tfrac{2}{3}K \gamma_{\alpha\beta}\chi- \tfrac{2}{3}K\omega_{(\alpha}u_{\beta)},
\eea
where     $\dot{\varpi}_{\alpha} \defn \big( \dot{ {\omega}}_{\alpha} - \tfrac{1}{3}K {\omega}_{\alpha}\big)$.
%  The kinetic and ``mass'' terms from the Einstein-Hilbert Lagrangian (\ref{eq:sec:mathcalt-m-einst-hilbert}) become
%\bea
%\mathcal{T} =\half \dot{\bar{h}}^{\mu\nu} \dot{\bar{h}}_{\mu\nu} - \half  \dot{\bar{h}} ^2+     \spnab^{\mu}\bar{h}{{}^{\nu}}_{\lambda}\spnab^{\lambda}\bar{h}_{\mu\nu}  - \half \spnab_{\lambda}\bar{h}^{\mu\nu} \spnab^{\lambda}\bar{h}_{\mu\nu}   -  \spnab^{\alpha}\bar{h}_{\alpha\lambda} \spnab^{\lambda}\bar{h} + \half  \spnab^{\lambda}\bar{h}  \spnab_{\lambda}\bar{h}  ,
%\eea
Substituting (\ref{eq:decomp-H-nudhfkdh-sec}) into the action (\ref{eq:sec:pert-met-only-ful_theory})  one   finds the absence of $\dot{\phi}^2$ and $\dot{n}^2_{\alpha}$ terms in the kinetic part of the Einstein-Hilbert Lagrangian; this can also be seen  in results given by \cite{Rubakov:2004eb, Rubakov:2008nh, Blas:2009my} and in the ADM formulation \cite{Arnowitt:1962hi}. $\phi$ and  $n_{\alpha}$ are now Lagrange multipliers whose equations of motion are constraint equations, allowing them to be eliminated. Using re-definitions of the coefficients, we can effectively set  $\phi=0$ and $n_{\alpha}=0$ which is equivalent to choice of the synchronous gauge. We will make this choice in what follows.

The two independent components of the equation of motion (\ref{eq:sec:xi-eom-vector}), after inserting the (3+1)-decomposition,  are given by
\bse
\label{eq:sec:eom_proj_full}
\bea
&& \big[\AW +\rho\big]\ddot{\chi} + \big[ \dot{A}_{\mathcal{W}}+\hct(4   \AW    +\rho        -   3P   )     \big]\dot{\chi}+  \big[P+\CW\big]\nabla^2\chi  \nonumber\\
&& -  \big[ \hct (3  \dot{P}+2  \dot{\rho}- \dot{A}_{\mathcal{W}}+3 \dot{B}_{\mathcal{W}})-  (2    \AW+5\rho +3P    -3    \BW -9    \DW  -6    \EW) \hct^2 \nonumber\\
&&\qquad\qquad+(2\rho+3     \BW  -     \AW) \tfrac{\ddot{a}}{a}+  \ddot{\rho}\big]\chi\nonumber\\
&&- \big[  \dot{B}_{\mathcal{W}} +\hct(3   \BW  +3   \DW  +2   \EW-2P)   \big]\partial_i\omega^i  + \big[\BW+\CW\big]\partial_i\dot{\omega}^i   \nonumber\\
&&\qquad\qquad =\frac{1}{2}   \big[\dot{B}_{\mathcal{W}}+\hct(3\BW+3 \DW  +2 \EW  -2P    )  \big]h+ \half\big[\BW-P\big]\dot{h}    ,
\eea
\bea
&&        \big[\rho-\CW\big]\ddot{\omega}^i - \big[     \dot{C}_{\mathcal{W}} +\hct(4\CW   -\rho          +    3  P)\big]\dot{\omega}^i -\big[\EW-P\big]\nabla^2\omega^i-  \big[\EW  + \DW\big] \partial^i\partial_k\omega^k\nonumber\\
&&\qquad + \big[  \dot{C}_{\mathcal{W}}-\hct(3 \DW  -\BW   +2 \EW   -4\CW-2P ) \big]\partial^i\chi +\big[\BW+\CW\big] \partial^i\dot{\chi}    \nonumber\\
&&\qquad\qquad   =
-\big[P-\EW\big] \partial_jh^{ij} +\half\big[P+\DW\big]\partial^ih .
\eea
\ese
$\rho$ and $P$ are the density and pressure coming from the dark fluid (i.e. the components of the background dark energy momentum tensor $U_{\mu\nu}$). The benefit of using the (3+1) decomposition has become apparent: we are able to identify the degrees of freedom.  There are the tensor degrees of freedom $h_+$ and $h_{\times}$ which are present in standard General Relativity, a vector degree of freedom $\omega^{i}$, that can be split into a longitudinal (scalar) and two transverse (vector) degrees of freedom, and a scalar degree of freedom $\chi$. Therefore, prima facie there are 6 extra degrees of freedom. As we will discuss below either $\chi$ or $\omega^{i}$ can be a ghost and therefore the coefficients must be chosen to suppress one or both of them. In the case where $\chi$ is the ghost then there are 5 degrees of freedom with those in $\omega^{i}$ being split into a longitudinal (scalar) and two transverse (vector) degrees of freedom. If $\omega^{i}$ is the ghost then there are only 3 degrees of freedom.

\section{Mechanisms for the elimination of ghosts} 

From (\ref{eq:action_s2_cons}) we see that the kinetic terms of the $\omega_{\mu}$ and $\chi$ fields  enter the theory via
\bea
\half \qsubrm{S}{\{2\}} &\supset& \int \dd^4x\sqrt{-g} \bigg[    (\AW + \rho)\dot{\chi}^2 + (\CW +P)\spnab_{\mu}\chi\spnab^{\mu}\chi \nonumber\\
&&\qquad\qquad\qquad + (\CW - \rho) \dot{\omega}_{\mu}\dot{\omega}^{\mu} +   (\DW+P)(\spnab_{\mu}\omega^{\mu})^2 + 2 (\EW-P) \spnab_{\mu}\omega_{\nu}\spnab^{\mu}\omega^{\nu}\bigg].
\eea
Let us now focus on the standard scenario of perturbations around Minkowski space-time when both $\rho = P=0$.
If $\AW >0$, then $\CW<0$ is required for $\chi$ to have a kinetic term with the ``proper sign''. But if this is the case then $\omega_{\mu}$  has a kinetic term with the ``wrong sign'' (the same is true if $\AW<0$). Hence, one of $\chi, \omega_{\mu}$ must be a ghost. There are a few ways to get out of this. 

First, one can make the coefficient of $\dot{\chi}^2$ vanish by setting $\AW =0$, which removes $\chi$ as a propagating mode and the equation of motion  is a constraint   that can be enforced in a Lorentz invariant theory. Secondly, one could set $\CW=0$, since that would remove $\omega_{\mu}$ as a propagating mode. Finally, one could set $\chi\equiv0$ directly which requires a breaking of Lorentz invariance (since we will be manually forcing one of the four components of a 4-vector to zero).

When $\AW=0$, there is no $\dot{\chi}^2$ term in the Lagrangian and the equation of motion simply becomes a constraint equation specifying the value of $\chi$ from the other field variables. This can be back-substituted into the action so that the theory explicitly does not contain the $\chi$-field.
From our presentation it is clear that in this case the ghost can be identified with the time-like degree of freedom $\chi$. We have been able to deduce this   since we used a (3+1) decomposition. Performing a transverse-longitudinal decomposition does not aid the identification of the ghost.  

If we choose the 5 parameters in (\ref{eq:Sec:31_w}) to be given by 
$\AW = X+Y$, $\BW = - X$, $\CW = - \tfrac{1}{2} Y$, $\DW = X$, $\EW = \tfrac{1}{2} Y$,
then the theory is Lorentz invariant and the mass-matrix is be given by
\bea
\label{eq:sec:sol-ghosyt-prot-w}
\mathcal{W}^{\mu\nu\alpha\beta} =  X g^{\mu\nu}g^{\alpha\beta} + Y g^{\mu(\alpha}g^{\beta)\nu},
\eea
where $X,Y$ are two parameters that are dependent on background field variables only. The standard route for isolating the ghost in the Lorentz invariant theory \cite{ArkaniHamed:2002sp} decomposes   $\xi^{\mu}$  into its transverse and longitudinal modes as
\bea
\label{eq:sec;decomps-xi}
\xi_{\mu}=\zeta_{\mu} +  \nabla_{\mu} \kappa,
\eea
where $\nabla_{\mu}\zeta^{\mu}=0$ and $\kappa$ is a scalar field.
For Minkowski background spacetime, inserting (\ref{eq:sec;decomps-xi-whole-dg-h-xi}), (\ref{eq:sec;decomps-xi}) and the mass-matrix  (\ref{eq:sec:sol-ghosyt-prot-w}) into the Lagrangian (\ref{eq:sec:sol-mass-H-W}), whilst assuming that $X,Y$ are constants, yields 
\bea
\label{eq:sec:sol-ghosyt-prot}
\sol &=& \frac{1}{8}(Xh^2 + Yh^{\mu\nu}h_{\mu\nu}) + \half Yh^{\mu\nu}\partial_{(\mu}\zeta_{\nu)} + \half (Xh \square\kappa + Yh^{\mu\nu}\partial_{\mu}\partial_{\nu}\kappa) + \half Y\partial^{\mu}\zeta^{\nu}\partial_{(\mu}\zeta_{\nu)} \nonumber\\
&&\qquad\qquad+ Y \partial^{(\mu}\zeta^{\nu)} \partial_{\mu}\partial_{\nu}\kappa+ \half(X+Y) (\square\kappa)^2  . 
\eea
This expression has made the ghost  problem manifest in the Lorentz invariant language. The existence of the last term, $(X+Y)(\square\kappa)^2$, means that ghosts are inevitable (see e.g. \cite{deUrries:1995ty, deUrries:1998bi, Creminelli:2005qk}). The cure is to set $X=-Y$, which removes the problematic kinetic term, and leaves the Pauli-Feriz mass-term, $\sol\supset \subsm{\ld}{\rm{PF}} = h^2 - h^{\mu\nu}h_{\mu\nu}$. The parameter choice $X=-Y$ is called the \textit{Pauli-Feirz tuning}, and will render  massive gravitons  ghost-free at linearized order on Minkowski backgrounds. In this case $\AW = 0$, $\BW = - X$, $\CW = \tfrac{1}{2}X$, $\DW = X$, $\EW  = -\tfrac{1}{2}X$ which is a special case of the more general situation discussed earlier.
 
Rather than retain Lorentz invariance and be forced to use the Pauli-Feirz tuning to remove the ghost, it has been suggest that one can just fix the field  $u^{\mu}\xi_{\mu}=0$ which implies that $\chi=0$, removing it as a physical degree of freedom. Of course, this is not really a solution to the problem of the ghost, since we have just set the field to zero. However, as we will see in the next section that it is possible to impose a symmetry which is equivalent to this. The condition $u^{\mu}\xi_{\mu}=0$  imposes an interesting structure upon the fields in theory when we use the  transverse-longitudinal split language. Using this and  (\ref{eq:sec;decomps-xi}) implies that
$\dot{\kappa} = - u^{\mu}\zeta_{\mu}$
which can be differentiated  to yield $\ddot{\kappa} = - u^{\mu}\dot{\zeta}_{\mu}.$
This shows us that the $\kappa$-field (i.e. the longitudinal component of the $\xi^{\mu}$-field) does not propagate. Instead, the $\supsm{n}{\rm th}$ time derivative of $\kappa$ is replaced by the $\supsm{(n-1)}{\rm th}$ time derivative of $\zeta_{\mu}$. Evaluating the ``kinetic scalars'' (\ref{eq:sec:31_kin_scalars}) for the scalar $\kappa$, yields
\bse
\bea
\nabla^{\mu}\kappa\nabla_{\mu}\kappa = -u^{\mu}u^{\nu}\zeta_{\mu}\zeta_{\nu} +\spnab^{\mu}\kappa\spnab_{\mu}\kappa,
\eea
\bea
\label{eq:sec:pi-after-simpl}
\square\kappa = u^{\mu}\dot{\zeta}_{\mu} + \spnab^{\mu}\spnab_{\mu}\kappa.
\eea
\ese
Using (\ref{eq:sec:pi-after-simpl}), the previously offensive term in (\ref{eq:sec:sol-ghosyt-prot}) becomes
\bea
(X+Y) (\square\kappa  )^2 &=& (X+Y) (u^{\mu}u^{\nu} \dot{\zeta}_{\mu} \dot{\zeta}_{\nu}  +2u^{\alpha}\dot{\zeta}_{\alpha} \spnab^{\mu}\spnab_{\mu}\kappa+   \spnab^{\mu}\spnab_{\mu}\kappa \spnab^{\alpha}\spnab_{\alpha}\kappa ), 
\eea
and we observe that the multiple derivatives of $\kappa$ that are present are entirely spatial. The upshot is that there are no time-derivatives of the scalar $\kappa$ left, and, \textit{crucially} no second time-derivatives of $\kappa$ in $\square\kappa$. The term $(X+Y)(\square\kappa)^2$ in (\ref{eq:sec:sol-ghosyt-prot}) is now longer problematic, and  does not require removal.

\section{Imposing reparameterization invariance}
A key aspect of the theories under consideration here is the spontaneous violation of reparameterization invariance. It is interesting to see under what conditions it can be be reimposed on the theory. Therefore, we consider how the vector field $\xi^{\mu}$  sources the perturbed gravitational field equations, and under what circumstances its components decouples from the field equations. From (\ref{eq:s2_action})   the field equations for the metric are $\ep G^{\mu\nu} = 8 \pi G \ep T^{\mu\nu} + \ep U^{\mu\nu}$, where $\ep U^{\mu\nu}$ is the dark energy momentum tensor and contains contributions to the field equations from the dark sector. In \cite{Battye:2012eu} we showed that 
\bea
\ep U^{\mu\nu} = - \half \big( \mathcal{W}^{\mu\nu\alpha\beta} + U^{\mu\nu}g^{\alpha\beta} \big) \delta g_{\alpha\beta}- \xi^{\alpha}\nabla_{\alpha}U^{\mu\nu} + 2U^{\alpha(\mu}\nabla_{\alpha}\xi^{\nu)}.
\eea
It is useful to note the terms in $\ep U^{\mu\nu}$ which are present due to the background dark energy-momentum tensor $U^{\mu\nu}$.
The components of $\ep U^{\mu\nu}$  are written as perturbed fluid variables,
\bea
\ep {U^{\mu}}_{\nu} = \delta \rho u^{\mu}u_{\nu} + 2 (\rho+P)v^{(\mu}u_{\nu)} + \delta P{\gamma^{\mu}}_{\nu} + P {\Pi^{\mu}}_{\nu},
\eea
where one can obtain
\bse
\label{eq:sec:met-only-xi0in}
\bea
\delta\rho &=& \bigg[\dot{\rho} +  \hct  \left(2\rho - \AW +3 \BW \right) \bigg]\chi- (\AW+\rho) \dot{\chi}  +(\rho + \BW)  \left(\half h  +\partial_i\omega^i\right) ,
\eea
\bea
 \delta P &=&  -\bigg[      \dot{P}   +    \hct   \left(2 P-\BW +3\DW +2\EW \right) \bigg]\chi + (\BW-P)\dot{\chi}       -\tfrac{1}{3}(P+3\DW+2\EW)  \left(  \frac{1}{2}h +   \partial_i\omega^i\right)   , 
\eea
\bea
(\rho+P)v^i =(P+\CW) \partial^i\chi + (\rho- \CW)   \dot{\omega}^i ,
\eea
\bea
P{\Pi^i}_j = 2(P-\EW)  \left( \half {h^i}_j+ \partial^{(i}\omega_{j)}- \tfrac{1}{3}{\delta^i}_j (\tfrac{1}{2}h + \partial_k\omega^k)\right) .
\eea
\ese
These effective fluid variables define how the components of the vector field $\xi^{\mu}$ sources the gravitational field equations. If one, or both, of the fields $\chi$ and $\omega^{i}$ does not appear in (\ref{eq:sec:met-only-xi0in}) then that field is not dynamical and hence can be completely ignored. It is clear that particular choices of the free functions in the mass-matrix it will be possible to achieve this. When one or both does not appear, it means that the theory is invariant under the symmetry associated with that field. Therefore, we can impose reparameterization invariance in three natural ways:

\begin{itemize}

\item  $\xi^{\mu}$-field decouples from the system when
$\dot{\rho} + 3\hct(\rho+P)=0$, $\dot{P} + \hct(P + 3\DW + 2 \EW)=0$, 
$\AW = - \rho$,  $\BW = P$,  $\CW = -P$, $\rho = -\BW$,  $\rho = \CW$, $P = \EW$,  $\DW = -P$. Hence, in the ``fully'' reparameterization invariant case, where the theory is forced to be invariant under $x^{\mu}\rightarrow x^{\mu}+\xi^{\mu}$,  the only values of $\rho, P$ that are allowed are those which are provided by a cosmological constant, $\rho = -P$, and all perturbed fluid variables vanish. Neither the $\chi$-  nor  the $\omega^{i}$-field propagate.
 
\item  $\chi = u_{\mu}\xi^{\mu} $ field decouples from the system when the parameters satisfy $\AW = - \rho$, $\BW = P$, $\CW = -P$, 
$\dot{\rho} +  \hct  \left(2\rho - \AW +3 \BW \right)=0$, $\dot{P}   +   \hct\left(2 P-\BW +3\DW +2\EW \right)=0$ from which we can deduce that  $\dot{\rho} + 3\hct(\rho+P)=0$ and $\dot{P} + \hct(P + 3\DW + 2 \EW)=0$. These equations appear to leave two coefficients, $\DW$ and $\EW$, unspecified. If we now define two parameters $\beta$ and $\mu$ via $\DW = \beta - P - \tfrac{2}{3}\mu$ and $\EW = \mu+P$ then we find that $\beta=(\rho+P)\textstyle{\dd P\over \dd\rho}$ which is the definition of the relativistic bulk modulus and $\mu$ can then be interpreted as a rigidity modulus of an elastic medium. Hence, in the case where we impose  time translation invariance, $t\rightarrow t+\chi$, but not spatial translation invariance, then we find that the theory must be EDE.  The equations of motion (\ref{eq:sec:eom_proj_full}) become
\bse
\bea
\label{eq:sec:chi-deoup_term}
&&  -3\hct\big[\dot{P} + 3 \beta\hct\big]\chi=0 ,
\eea
\bea
\label{eq:sec:chi-deoup_term-b}
&&        \big[\rho+P\big]\ddot{\omega}^i + \big[   \dot{P} +\hct(P   +\rho )\big]\dot{\omega}^i-    \big[\EW-P\big]\nabla^2\omega^i-  \big[\EW  + \DW\big] \partial^i\partial_k\omega^k\nonumber\\
&&     \qquad\qquad =-\big[P-\EW\big] \partial_jh^{ij} +\tfrac{1}{2}\big[P+\DW\big]\partial^ih .
\eea
\ese
Note that (\ref{eq:sec:chi-deoup_term}) vanishes for arbitrary values of $\chi$ since $\beta = (\rho+P)\dot{P}/\dot{\rho}$, and that there is a propagating vector degree of freedom, $\omega^i$. (\ref{eq:sec:chi-deoup_term-b}) is the equation of motion presented in \cite{PhysRevD.76.023005}. In this case the mass-term for the gravitons is
\bea
\sol = \tfrac{1}{8}\rho \bigg[ (w^2 - \tfrac{2}{3}\hat{\mu})\bar{h}^2 + 2 (w+\hat{\mu})\bar{h}_{\mu\nu}\bar{h}^{\mu\nu}\bigg],
\eea
where $w = P/\rho, \hat{\mu} \defn \mu /\rho$. This case is equivalent to setting $\AW=0$ in the Minkowski space case. Since $\chi$ does not appear in (\ref{eq:sec:met-only-xi0in}) it is no longer a physical degree and there is no ghost.

\item $\omega^i$ decouples when $\rho + \BW = 0$,  $P+3\DW+2\EW=0$, $\rho = \CW$, $P = \EW$ from which we can deduce that $\BW = - \CW =- \rho$ and $\DW = - \EW = -P$.  Therefore, we see that in the case where we impose spatial translation invariance, $x^i\rightarrow x^{i}+\xi^{i}$, but not time translation invariance, then the perturbations have some of the characteristics of massive scalar field theory as explained in \cite{Battye:2012eu}. The equations of motion (\ref{eq:sec:eom_proj_full}) become
 \bse
\bea
&& \big[\AW +\rho\big]\ddot{\chi} + \big[ \dot{A}_{\mathcal{W}}+\hct(4   \AW    +\rho        -   3P   )     \big]\dot{\chi}+  \big[\rho+P\big]\nabla^2\chi  \nonumber\\
&&\qquad\qquad +  \big[(\dot{A}_{\mathcal{W}} + 3 (\AW - \rho-2P)\hct)\hct + (\AW + 4\rho + 3 P)\dot{\hct}  \big]\chi  = - \tfrac{1}{2}\big[ \rho+P\big]\dot{h}    ,
\eea
\bea
\label{eq:sec:chi-deoup_term-2b}
&&      \big[ \dot{\rho} + 3 \hct(\rho+P)\big]\partial^i\chi      =0 .
\eea
\ese
Note that (\ref{eq:sec:chi-deoup_term-2b}) vanishes for arbitrary values of $\chi$ due to the background conservation equation. If we define the entropy
\bea
w\Gamma=\left({\delta P\over \delta\rho}-w\right)\delta\,,
\eea
and set $\AW=[w+\epsilon(1+w)]\rho$ we find that 
\bea
w\Gamma=\left({1\over 1+\epsilon}-w\right)\left[\delta - 3\hct\left( {w(1+\epsilon)+1\over w(1+\epsilon)-1}\right)(1+w)\theta\right]\,.
\eea
This implies that this theory is a scalar field theory with a non-standard kinetic term. The mass-term is given by
\bea
\sol = - \tfrac{1}{8}w\rho\bigg[\bar{h}^2 - 2\bar{h}_{\mu\nu}\bar{h}^{\mu\nu}\bigg],
\eea
which we note does not satisfy the Pauli-Feirz tuning. This case is equivalent to setting $\CW=0$ in the Minkowski space case. Since $\omega^{i}$ does not appear in (\ref{eq:sec:met-only-xi0in}) it is no longer a physical degree and there is no ghost.
\end{itemize}

\section{Discussion}
It is well established in the literature, that can be   ghosts in \textit{general} massive gravity theories, and various methods have been devised to remove them.  If one imposes Lorentz invariance, then one is forced to use the Pauli-Feirz tuning to excise the ghost. If one is willing to give up Lorentz invariance, then certain parameter choices allow for ghost-free massive gravity theories. We have shown that the theory (\ref{eq:sec:sol-mass-H-W}) with the (3+1)-decomposition of the mass-matrix (\ref{eq:Sec:31_w}) imposed with just time-translation invariance constitutes a linearized theory with  healthy Lorentz-violating massive gravitons with 5 physical degrees of freedom. That theory \textit{is} exactly the EDE model previously discussed in the literature \cite{PhysRevD.76.023005}. In addition, if one imposes spatial translation invariance then there is another ghost-free massive gravity theory with 3 degrees of freedom.

In terms of the EDE parameters and graviton masses, $\AW = m_0^2 = - \rho,  \BW = - 2 m_4^2 = P,\CW = 4m_1^2 = - P, \DW = 4m_3^2 = \beta - P - \tfrac{2}{3}\mu, \EW = - 2 m_2^2 = \mu +P$.
The masses can be conveniently parameterized by some overall mass scale $M^2 \defn \rho\sim H_0^2$ (since we wrote $U_{\mu\nu} = \rho u_{\mu}u_{\nu} + P\gamma_{\mu\nu}$, $\rho$ has units of mass squared),
\bea
\label{eq:sec:masses-ede}
&m_0^2 = - M^2,\quad m_1^2 =- \tfrac{1}{4}wM^2,\quad m_2^2 =- \tfrac{1}{2}(\hat{\mu}+w)M^2,\quad
& m_3^2 = \tfrac{1}{4}(w^2-\tfrac{2}{3}\hat{\mu})M^2,\quad m_4^2 = -\tfrac{1}{2}wM^2. 
\eea
We defined $\hat{\mu} \defn \mu/\rho$ in analogy with $w = P/\rho$.

\begin{figure*}[!t]
      \begin{center}
{\includegraphics[scale=1]{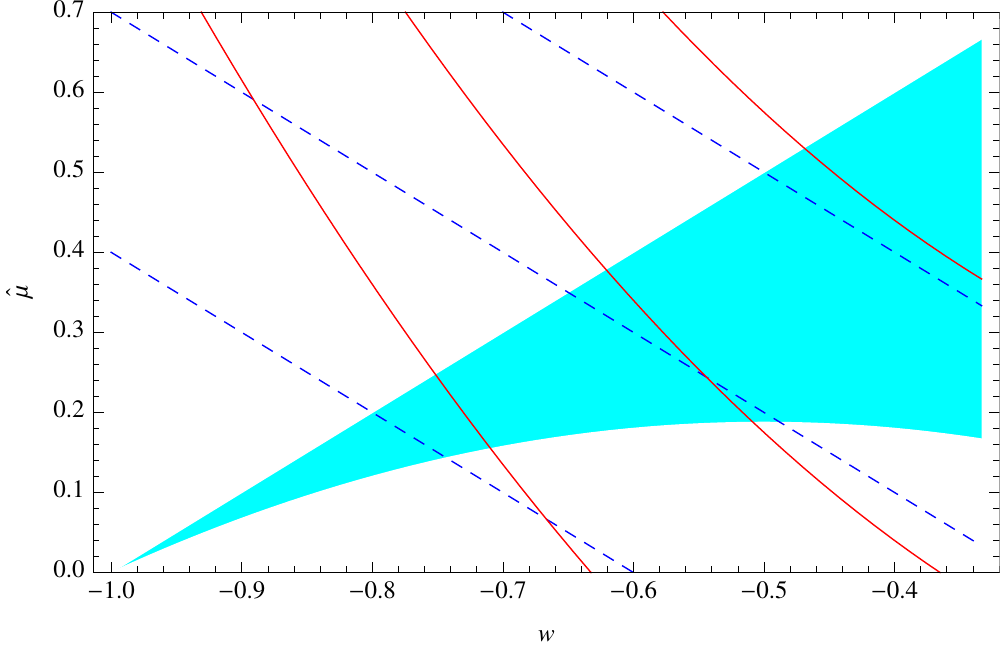}}
      \end{center}
\caption{The shaded region denotes the range of values of the equation of state $w$ and shear modulus $\hat{\mu}$ which yield sound speeds less than unity, which is where   the inequalities (\ref{eq:sec:allowes_mu}) are satisfied. The red (solid) lines denote lines of constant  $m_2^2 = \tfrac{1}{2}M^2(0.6,0.3,0)$ from left to right and the blue (dashed) lines of constant $m_3^2 = \tfrac{1}{6}M^2(0.6,0.2,-0.2)$, again from left to right. } \label{fig:allowedmasses}
\end{figure*}

To connect to dark energy, we note that the fraction of the total energy density that is dark energy is linked to the mass scale $M^2$ via $\qsubrm{\Omega}{de} = M^2/(3H_0^2)$.
Hence, we see that the ``natural" scale for the masses in order for the modification of gravity to act as a source of cosmic acceleration is of the order the Hubble parameter, and are all multiplied by order unity ``corrections'' defined by two parameters which encode the properties of the elastic medium: its equation of state parameter $w$ and shear modulus $\hat{\mu}$.
The longitudinal and transverse sound speeds of EDE are \cite{PhysRevD.76.023005}
\bea
\qsubrm{c}{s}^2 = w + \frac{\tfrac{4}{3}\hat{\mu}}{1+w},\qquad \qsubrm{c}{v}^2 = \frac{\hat{\mu}}{1+w}.
\eea
Stability and subluminality require that $0\leq \qsubrm{c}{i}^2 \leq 1$, so that we   have the following constraints on the possible values of $\hat{\mu}$
\bea
\label{eq:sec:allowes_mu}
- \tfrac{3}{4}w(1+w) \leq \hat{\mu} \leq \tfrac{3}{4}(1-w^2),\qquad 0 \leq \hat{\mu} \leq 1+w.
\eea
In \fref{fig:allowedmasses} we plot the allowed values of $(w,\hat{\mu})$ which satisfy (\ref{eq:sec:allowes_mu}), and some lines of constant $m_2^2$ and $m_3^2$. Observationally   the values of $w, M^2$ and $\hat{\mu}$ (only $\hat{\mu}$   is the ``new'' parameter) can be constrained and then   (\ref{eq:sec:masses-ede}) used to obtain the graviton masses.

Elastic dark energy and massive gravity share two common features. First, they are both constructed from rank-4 tensors, (the elasticity tensor and mass-matrix, respectively) and these tensors have identical symmetries in their indices. Secondly, they both have five propagating degrees of freedom. The extra degrees of freedom in elastic dark energy may have a different fundamental origin to those in massive gravity, but they enable an interesting interpretation to be extracted from massive gravities. Our interpretation is that \textit{massive gravity is the manifestation  of rigidity of spacetime}.

\textit{\textbf{Acknowledgements}}
We have appreciated  conversations with Niayesh Afshordi, Stephen Appleby, Ruth Gregory and Kurt Hinterbichler.  JAP  is supported by the STFC Consolidated Grant ST/J000426/1. This research was supported in part by Perimeter Institute for Theoretical Physics. Research at Perimeter Institute is supported by the Government of Canada through Industry Canada and by the Province of Ontario through the Ministry of Economic Development and Innovation.
\clearpage

%\bibliographystyle{unsrt}
%\bibliographystyle{JHEP}
%\bibliography{refs,selfrefs}
\providecommand{\href}[2]{#2}\begingroup\raggedright\endgroup

\end{document}